\documentclass[aps,prb,notitlepage,a4paper,floatfix,nofootinbib]{revtex4-1}

\usepackage{graphicx}
\usepackage{amsmath}

\newcommand{\Dp}{\Delta p}
\newcommand{\eg}{{\it e.g.}, }

\newcommand{\ie}{{\it i.e.}, }
\newcommand{\kB}{k_{\rm B}}
\newcommand{\muw}{\mu_{\rm w}}
\newcommand{\Ns}{N_{\rm s}}

\newcommand{\pin}{p_{\rm in}}
\newcommand{\po}{p_{\rm out}}
\newcommand{\Qc}{Q_{\rm c}}
\newcommand{\tM}{\tilde{M}}

\begin{document}

\title{Vesicles as osmotically stressed capsules}

\author{Emir Haleva}
\affiliation{Raymond \& Beverly Sackler School of Chemistry, Tel Aviv University,
Tel Aviv 69978, Israel}
\author{Haim Diamant}
\email{hdiamant@tau.ac.il}
\affiliation{Raymond \& Beverly Sackler School of Chemistry, Tel Aviv University,
Tel Aviv 69978, Israel}

\begin{abstract}
  Vesicular capsules are used to carry biochemicals in biology and
  liposome technology. Being water-permeable with differing interior
  and exterior compositions, they are necessarily under osmotic
  stress. Recent studies have underlined the different thermodynamic
  behavior of osmotically stressed vesicles in comparison to vesicles
  subjected to a hydrostatic pressure as studied earlier. Through
  their different behavior one gains access to the parameters
  affecting the osmotic swelling of vesicles, such as the
  membrane-permeability coefficients of solute molecules.
\end{abstract}

\maketitle

\section*{Introduction}

Bilayer vesicles serve as capsules for delivery of various molecules
as part of the biological cell function \cite{bio} and in
pharmaceutical and cosmetic applications.\cite{Lasic} This utility
relies on the potential barrier posed by the hydrophobic core of the
bilayer for the penetration of water-soluble molecules. The
sensitivity of molecular transport to the barrier height makes the
membrane-permeability coefficients of different solutes span many
orders of magnitude\,---\,from values of order $10^2$ $\mu$m/s for
water down to a mere $\sim 10^{-8}$ $\mu$m/s for potassium ions, for
example.\cite{Paula1996} Thus, the membrane of a vesicle of micron
size or smaller, over time scales longer than about $10^{-2}$ s,
behaves as a semi-permeable partition, allowing solvent to be
exchanged between the interior and exterior while keeping certain
solute molecules either enclosed inside or locked outside. This, in
turn, allows for the buildup of an osmotic pressure difference across
the membrane.

We open with a very basic question: What are the thermodynamic
constraints imposed in practice on such a vesicular capsule? The
surrounding solution dictates the temperature $T$, external pressure
$\po$, and chemical potential $\muw$ of the solvent (water). Over the
time scales under consideration the numbers $\{Q_i\}$ of the
encapsulated solute molecules are fixed as well. On the other hand,
since solvent is exchanged across the membrane, the values of neither
the encapsulated volume $V$ nor the inner pressure $\pin$ are {\it a
priori} set. In addition, there are surface constraints associated
with the membrane, such as the number $\Ns$ of amphiphilic molecules
making the bilayer (or, alternatively, their chemical potential
$\mu_{\rm s}$).  These two-dimensional variables raise various subtle
issues,\cite{pre11} which nonetheless do not concern us here; we
simply represent them all by a single symbol, $\sigma$. The set of
thermodynamic constraints imposed on the capsule, therefore, is
$(T,\po,\muw,\{Q_i\},\sigma)$.

Theoretically, vesicles have been studied over the years under
different sets of constraints, such as $(T,V,\sigma)$
\cite{Helfrich1973,Seifert1997} or
$(T,\Dp=\pin-\po,\sigma)$.\cite{GompperKroll} The former will hold in
practice when the solvent does not have sufficient time to be
exchanged across the membrane, \ie over sufficiently short time scales
(shorter than $\sim 10^{-2}$ s). The latter corresponds, for example,
to micropipette aspiration experiments,\cite{Evans1990} where the
hydrostatic pressure difference across the membrane is controlled.

The key point that we wish to highlight here is that the behaviors of
vesicles under these different sets of thermodynamic constraints are
not necessarily equivalent. This has been demonstrated in a series of
recent studies.\cite{epje06b,pre08,prl08,sm12,advances12} We are used
to the fact that the distinction between different sets of
thermodynamic constraints (equivalently, different statistical
ensembles) is not important for large systems at equilibrium. Why
should it matter much whether a certain pressure difference is
externally imposed by a pump or self-attained as an average
equilibrium value due to osmosis?  This broadly valid statement
assumes, however, that surface effects and fluctuations are
negligible. For fluid vesicles, which often strongly fluctuate and may
introduce strong surface effects, the validity of this assumption is
not self-evident.  It should be stressed that the distinction between
sets of constraints is not merely a theoretical curiosity. The
explicit treatment of the encapsulated solution has in certain cases
important and useful implications. We choose to postpone the
discussion of these to the end of this Highlight piece and discuss
first two much more artificial cases, which nevertheless serve well to
demonstrate the point.

\section*{Osmotic swelling of model capsules}

Arguably the simplest model for a fluctuating closed envelope is a
two-dimensional (2D) ring made of a fixed number $\Ns$ of
freely-jointed segments at temperature $T$. (Such systems are actually
realizable experimentally.\cite{Severin2006}) When we control and
increase the 2D pressure difference $\Dp$ between the inner and outer
regions, the mean volume enclosed by the envelope (\ie the area of the
ring) increases. For self-intersecting rings this swelling encounters
a criticality\,---\,at a critical pressure, $\Dp=\Dp_{\rm c}\propto
\Ns^{-1}$, the mean volume either diverges (if the segments are taken to
be extensible harmonic springs) \cite{RudnickGaspari} or exhibits a
second-order transition between crumpled and smooth states (if the
segments are made inextensible).\cite{epje2006a} However, when we let
the ring swell due to an increasing number $Q$ of enclosed particles
rather than impose a pressure difference, the criticality disappears
and the mean volume increases gradually with $Q$.\cite{epje06b}

A similar clear-cut example of a qualitatively different behavior of
particle-encapsulating envelopes is found in a discrete model of
three-dimensional (3D) fluid vesicles. In this model, due to Gompper
and Kroll,\cite{GompperKroll} the vesicle is represented by a closed
triangulated network of $\Ns$ self-avoiding nodes whose connectivity
is random and variable. When the pressure difference $\Dp$ is
controlled and increased, the vesicle undergoes a first-order
transition at a certain critical pressure, $\Dp=\Dp_{\rm c}\propto
\Ns^{-1/2}$, between crumpled and smooth states. However, if the
vesicle is inflated instead by an increasing number $Q$ of enclosed
particles, the discontinuous transition disappears and is replaced by
gradual swelling with $Q$.\cite{pre08}

\begin{figure}[tbh]
\centerline{\resizebox{0.5\textwidth}{!}{\includegraphics{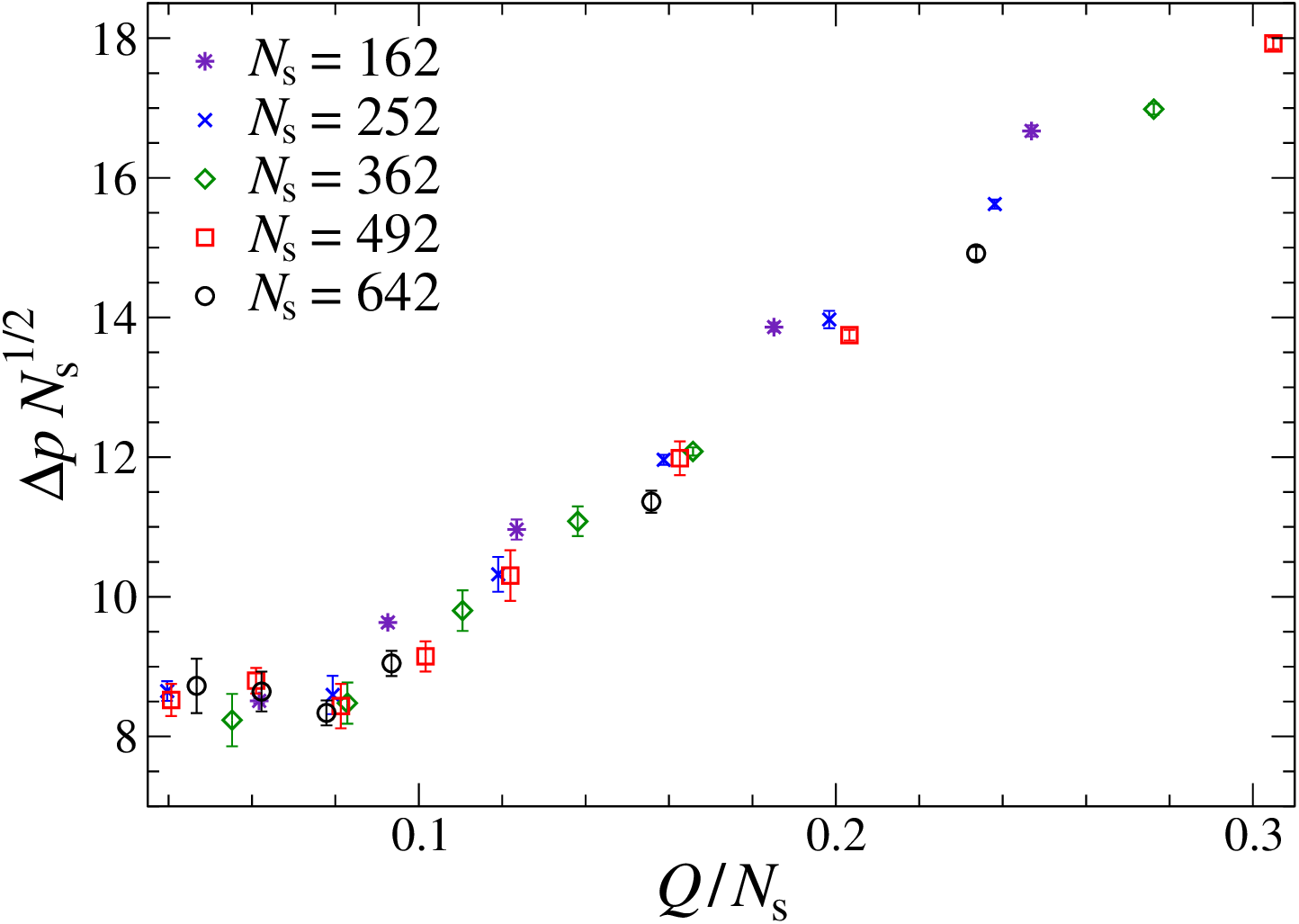}}}
\caption{Mean pressure difference across the membrane of a vesicle as
  a function of encapsulated particle number. As $Q$ decreases, so
  does the mean volume, and $\Dp$ never reaches arbitrarily small
  values. Data were obtained from Monte Carlo simulations of the
  Gompper-Kroll fluid vesicle model for several vesicle sizes (number
  of surface nodes $\Ns$) as indicated. The pressure was calculated
  from the mean particle concentration $c$ assuming an ideal solution,
  $\Dp = \kB T c$. Axes are rescaled by the appropriate power of
  vesicle size to achieve data collapse. For more details see
  ref~\citenum{pre08}.}
\label{fig_Dp}
\end{figure}

The way in which the criticality is removed in these two examples is
quite unusual. When we control $Q$, neither $\Dp$ nor $V$ is
fixed. The large flexibility of those two model capsules allows them
to adjust their mean volume such that the mean pressure difference,
determined by the mean particle concentration $Q/V$ and $T$, never
hits $\Dp_{\rm c}$ for any value of $Q$, thus avoiding the
transition. This is demonstrated in Fig.\ \ref{fig_Dp}. Upon
decreasing the number of enclosed particles the pressure inside the 3D
vesicle first decreases but eventually stops changing with $Q$ and
never reaches arbitrarily low values. Thus, the two sets of
constraints, $(T,\Dp,N)$ and $(T,\po,Q,N)$, are manifestly not
equivalent in these examples\,---\,there are macrostates in the former
that become inaccessible in the latter. (For a unified description of
the swelling of random manifolds by either imposed pressure or
enclosed particles, and the related issue of equivalence, see
ref~\citenum{pre08}.)

\section*{Osmotic swelling of membrane vesicles}

The two examples given in the preceding section pertain to random
envelopes whose size is much larger than their bending persistence
length.\cite{Lipowsky1991} Actual bilayer vesicles belong to the opposite
smooth limit, where the bending persistence length is larger than the
vesicle. When real vesicles swell, they first undergo an ``ironing''
stage, where their volume-to-area ratio increases as they become
increasingly more spherical. This is followed by a stretching stage,
in which the membrane area $A$ increases while the shape remains
essentially spherical. When the surface strain exceeds a few percent,
the membrane ruptures.\cite{Evans1990,sm12} In the case of osmotic swelling,
the rupture (osmotic lysis) may be followed by a sequence of
additional swelling--rupture cycles.\cite{Peterlin2008}

We focus here on the crossover between the ironing and stretching
stages. We define an order parameter, 
\begin{equation}
  M = 1 - 6\sqrt{\pi}V/A^{3/2},
\label{orderparameter}
\end{equation}
which measures how far the vesicle is from a sphere.  Way before the
crossover, when the vesicle is not swollen, $M$ is appreciable, whereas
well into the stretching stage, when the vesicle is nearly spherical,
$M$ approaches zero. It was found out that, in the case of osmotic
swelling, this crossover can be represented as a rounded continuous
phase transition.\cite{prl08,sm12} Specifically, if a control
parameter $q$, dependent on the number of encapsulated particles, is
defined as
\begin{equation}
  q = Q/\Qc - 1,\ \ 
  \Qc = \po V_0 / (\kB T),\ \ 
  V_0 = A_0^{3/2}/(6\sqrt{\pi}),
\label{controlparameter}
\end{equation}
$A_0$ being the relaxed area of the vesicle, one finds the following
critical scaling in the vicinity of the transition and beyond it:
\begin{equation}
  M = \Delta \tM(q/\Delta).
\label{scaling}
\end{equation}
In eqn (\ref{scaling}) $\Delta$ is the width of the transition, and
$\tM(x)=(\sqrt{1+x^2}-x)/2$ is a scaling function. (The expressions
given above for $\Qc$ and $\tM$ assume that the solution inside the
vesicle is ideal; yet, the results can be generalized to an arbitrary
equation of state for the encapsulated solution.\cite{prl08})

In the limit $\Delta\rightarrow 0$ the swelling curve $M(q)$ has a
singular corner at the point $(q=0,M=0)$.  If membrane stretching is
neglected (the area being fixed at $A=A_0$), then $\Delta\propto
\Ns^{-1/4}$\,---\,\ie the rounding of the transition arises, as in any
phase transition, from the finite size of the system.\cite{prl08} In
this approximation there is no stretching stage, and the vesicle
approaches its maximum volume $V_0$, the volume of a sphere of area
$A_0$. When a finite stretching modulus $K$ is included, the rounding
of the transition arises from both finite size and finite
stretchability (with $\Delta\propto K^{-1/2}$ if the latter dominates
\cite{sm12}).

The critical properties of the crossover disappear when the swelling
is hydrostatic rather than osmotic, \ie when the pressure difference
is constrained instead of the number of enclosed solute
molecules. Unlike the examples in the preceding section, in this case
the same macrostates are encountered when sweeping through the values
of either $\Dp>0$ or $Q$; it is the sharpness of the corresponding
changes in the order parameter that differs between the two sets of
constraints. When the vesicle swells by osmosis, {\em both} the mean
volume and mean pressure behave critically as a function of $Q$, and
we find $\Dp(q)\propto 1/M(q)$;\cite{prl08,sm12} the sharp approach to
a spherical shape is accompanied by a similarly sharp increase in
pressure difference and surface tension. Consequently, once
transformed into pressure dependence, the order parameter decreases
slowly with increasing pressure, $M(\Dp)\propto 1/\Dp$.

The critical scaling has had two beneficial implications. One is
theoretical\,---\,eqn (\ref{scaling}) constitutes a law of
corresponding states for the osmotic swelling of vesicles.\cite{sm12}
It implies that the osmotic swelling curves of various vesicles under
various conditions (\eg due to the permeation of various solutes at
various concentrations) can be collapsed in the vicinity of the
transition (and above it) onto a single master curve, thus achieving a
simple, unified theoretical description.

\begin{figure}[tbh]
\centerline{\resizebox{0.5\textwidth}{!}{\includegraphics{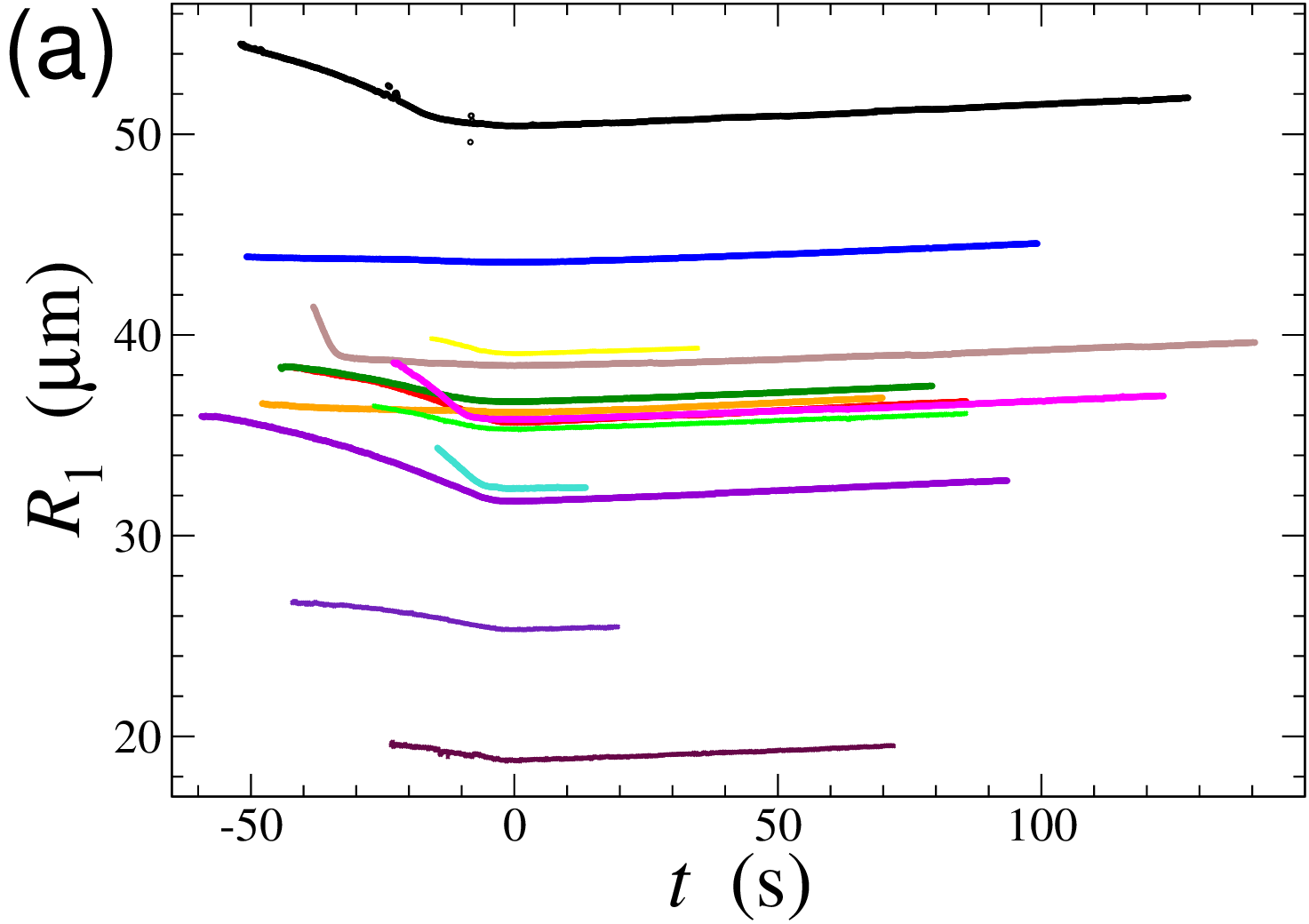}}}
\vspace{1cm}
\centerline{\resizebox{0.5\textwidth}{!}{\includegraphics{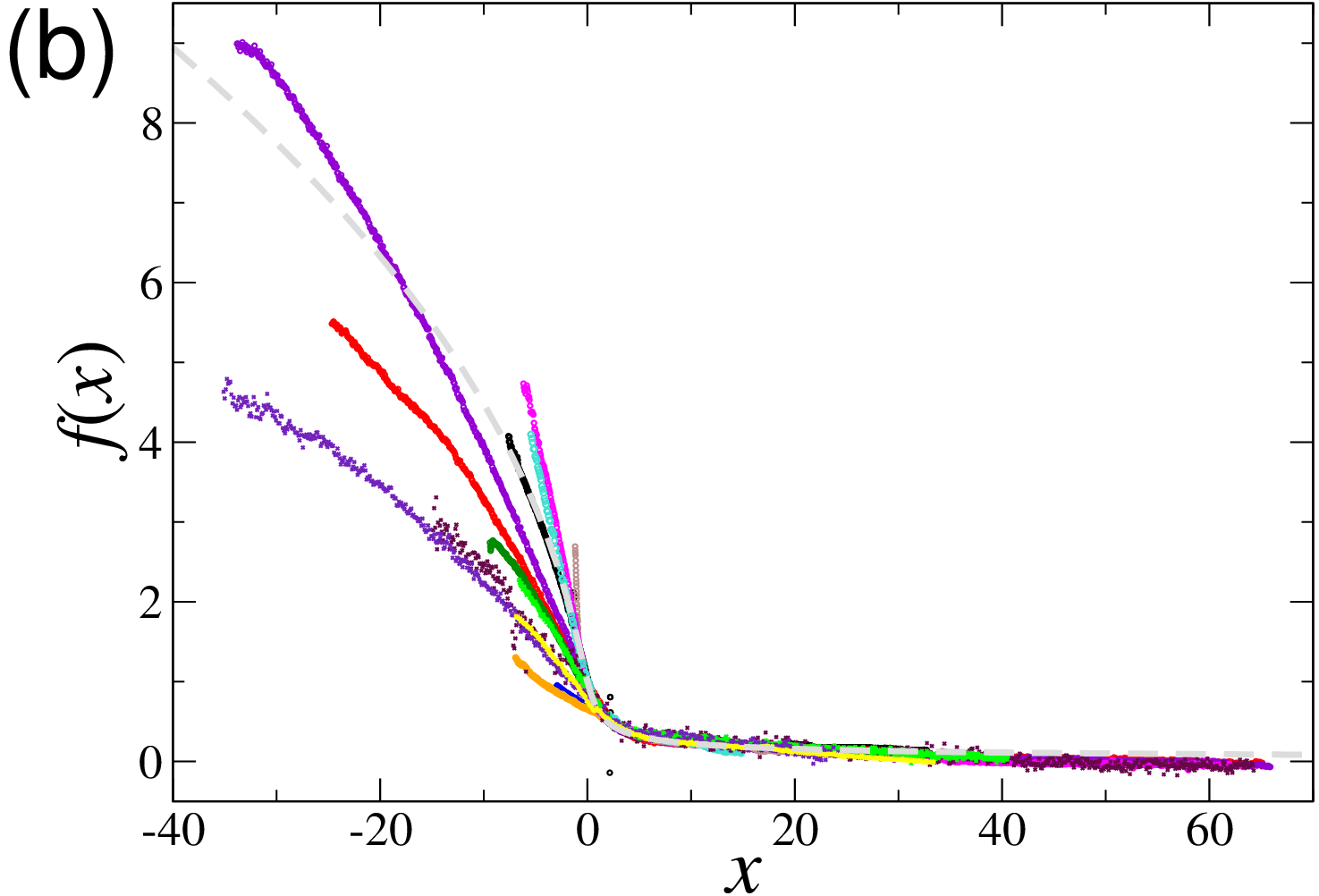}}}
\caption{(a) Osmotic swelling of POPC giant unilamellar vesicles due
  to the permeation of urea. Each curve shows the optically tracked
  principal radius of a spheriodal vesicle as a function of time. The
  decreasing part corresponds to the ironing stage, and the increasing
  part to the stretching stage, after which the vesicle ruptures. The
  curves are flattened because of the large polydispersity of
  vesicles. (b) Collapse of the swelling curves onto the master curve,
  $f(x)=(\sqrt{1+x^2}-x)^{1/2}$ (represented by a gray dashed
  line). For more details, including the precise rescaling scheme, see
  ref~\citenum{sm12}.}
\label{fig_collapse}
\end{figure}

Osmotic swelling has been experimentally studied in detail using
dynamic light scattering \cite{Paula1996} or optical
tracking,\cite{Peterlin2008} depending on vesicle size. In a typical
procedure vesicles are formed in a solution of a non-permeating
solute, in which they are free of osmotic stress. Subsequently, the
exterior is replaced by a solution of equal concentration but
containing a permeating solute. As the outer solute permeates inward,
the vesicles swell through osmosis. Changes in the shape and size of
the vesicle as it progresses from a stress-free state toward a
strongly stretched sphere and the ultimate lysis are related to the
corresponding decrease in the order parameter $M$, eqn
(\ref{orderparameter}). The control parameter $q$, eqn
(\ref{controlparameter}), is linearly related to the time axis via the
membrane permeability coefficient, $P$, of the permeating
solute. Thus, high-resolution optical tracking can be used to obtain
swelling curves, $M(q)$, and the associated data collapse.\cite{sm12}
This is demonstrated in Fig.\ \ref{fig_collapse}.

The other beneficial implication of the critical scaling has been the
introduction of a new method to measure membrane-permeability
coefficients of various solutes.\cite{sm12,advances12} Data collapse
such as the one shown in Fig.\ \ref{fig_collapse} involves two fitting
parameters: the transition width $\Delta$, separately fitted for each
curve, and the permeability coefficient $P$, globally fitted for all
curves and corresponding to a mere scaling of the horizontal (time)
axis. This yields a sensitive and reliable measurement of $P$. For
example, from the data shown in Fig.\ \ref{fig_collapse} one finds for
the permeation of urea through a POPC membrane $P=0.013\pm 0.001$
$\mu$m/s. A similar procedure has yielded a significant concentration
dependence of $P$ for polyols (glycerol and ethylene glycol), which
was not recognized before.\cite{sm12,advances12}

\section*{Conclusion}

The examples given above demonstrate how useful vesicular capsules may
be for investigating fundamental issues of the thermodynamics of small
systems, such as strong surface effects, fluctuations, and ensemble
equivalence. The different behavior of osmotically stressed vesicles,
as highlighted here, makes it necessary in certain cases to explicitly
consider the properties of the encapsulated solution rather than just
specify its mean volume or pressure. At the same time the different
behavior allows access to features of the osmotic swelling process
that would otherwise be difficult to extract, such as the
membrane-permeability coefficients. One may be able to utilize it
further and come up with more detailed predictions\,---\,for example,
regarding the average time between the onset of osmotic stress and
membrane lysis, which may be important for drug delivery and release.

\section*{Acknowledgements}

We are grateful to Primo\v{z} Peterlin for a fruitful collaboration
that led to several of the insights presented here. Acknowledgment is
made to the Donors of the American Chemical Society Petroleum Research
Fund for partial support of this research (Grant No.\ 46748-AC6).

\end{document}